# Gyrification from constrained cortical expansion

Tuomas Tallinen[a], Jun Young Chung[b], John S. Biggins[c], and L. Mahadevan[d,e,f,1]

[a]Department of Physics and Nanoscience Center, University of Jyväskylä, FI-40014 Jyväskylä, Finland; [b]School of Engineering and Applied Sciences, Harvard University, Cambridge, MA 02138; [c]Cavendish Laboratory, Cambridge University, Cambridge CB3 0HE, United Kingdom; [d]Wyss Institute for Biologically Inspired Engineering, [e]Kavli Institute for Bionano Science and Technology, and [f]School of Engineering and Applied Sciences, Department of Organismic and Evolutionary Biology, and Department of Physics, Harvard University, Cambridge, MA 02138

The exterior of the mammalian brain—the cerebral cortex—has a conserved layered structure whose thickness varies little across species. However, selection pressures over evolutionary time scales have led to cortices that have a large surface area to volume ratio in some organisms, with the result that the brain is strongly convoluted into sulci and gyri. Here we show that the gyrification can arise as a nonlinear consequence of a simple mechanical instability driven by tangential expansion of the gray matter constrained by the white matter. A physical mimic of the process using a layered swelling gel captures the essence of the mechanism, and numerical simulations of the brain treated as a soft solid lead to the formation of cusped sulci and smooth gyri similar to those in the brain. The resulting gyrification patterns are a function of relative cortical expansion and relative thickness (compared with brain size), and are consistent with observations of a wide range of brains, ranging from smooth to highly convoluted. Furthermore, this dependence on two simple geometric parameters that characterize the brain also allows us to qualitatively explain how variations in these parameters lead to anatomical anomalies in such situations as polymicrogyria, pachygyria, and lissencephalia.

brain morphogenesis | elastic instability

The mammalian brain is functionally and anatomically complex. Over the years, accumulating evidence (1, 2) shows that there are strong anatomical correlates of its information-processing ability; indeed the iconic convoluted shape of the human brain is itself used as a symbol of its functional complexity. This convoluted (gyrified) shape is associated with the rapid expansion of the cerebral cortex. Understanding the evolutionary and developmental origins of the cortical expansion (1–6) and their mechanistic role in gyrification is thus an important question that needs to be answered to decipher the functional complexity of the brain.

Historically there have been three broad hypotheses about the origin of sulci and gyri. The first is that gyri rise above sulci by growing more (7), requiring the pattern of sulci and gyri to be laid down before the cortex folds, presumably by a chemical morphogen. There is no evidence for this mechanism. The second hypothesis considers that the outer gray matter consists of neurons, and the inner white matter is largely long thin axons that connect the neurons to each other and to other parts of the nervous system and proposes that these axons pull mechanically, drawing together highly interconnected regions of gray matter to form gyri (8–10). However, recent experimental evidence (11) shows that axonal tension when present is weak and arises deep in the white matter and is thus insufficient to explain the strongly deformed gyri and sulci. The third hypothesis is that the gray matter simply grows more than the white matter, an experimentally confirmed fact, leading to a mechanical buckling that shapes the cortex (11–14). Evidence for this hypothesis has recently been provided by observations of mechanical stresses in developing ferret brains (11), which were found to be in patterns irreconcilable with the axonal tension hypothesis. In addition, experiments show that sulci and gyri can be induced in usually smooth-brained mice by genetic manipulations that promote cortical expansion (15, 16), suggesting that gyrification results from an unregulated and unpatterned growth of the cortex relative to sublayers.

Nevertheless, there is as yet no explicit biologically and physically plausible model that can convincingly reproduce individual sulci and gyri, let alone the complex patterns of sulci and gyri found in the brain. Early attempts to mechanically model brain folding (13) were rooted in the physics of wrinkling and assumed a thin stiff layer of gray matter that grows relative to a thick soft substrate of white matter. This model falls short in two ways. First, the gray matter is neither thin nor stiff relative to the white matter (17, 18). Second, this model predicts smooth sinusoidal wrinkling patterns, sketched in Fig. 1A, whereas even lightly folded brains have smooth gyri but cusped sulci. More complicated mechanical models including, e.g., elasto-plasticity and stress-related growth (14, 19, 20), lead to varying morphologies, but all produced simple smooth convolutions rather than cusped sulci.

A fundamentally different mechanical instability that occurs on the surface of a uniformly compressed soft solid (21, 22) has recently been exposed and clarified, theoretically, computationally, and experimentally (23–26). This sulcification instability arises under sufficient compression leading to the folding of the soft surface to form cusped sulci via a strongly subcritical transition. In Fig. 1B, we show a geometry dual to that associated with wrinkling: A soft layer of gray matter grows on a stiff white-matter substrate. Unlike wrinkling, this instability can produce the cusped centers of sulci, but the flat bottom of the gray matter is not seen in the brain. This is a consequence of the assumption that the gray matter is much softer than the white matter—in reality the two have very similar stiffnesses (17, 18). We are thus led to the final simple alternate, sketched in Fig. 1C, where the stiffnesses of the gray and white matter are assumed to be identical. Such a system is subject to a cusp-forming sulcification

> **Significance**
>
> The convolutions of the human brain are a symbol of its functional complexity and correlated with its information-processing capacity. Conversely, loss of folds is correlated with loss of function. But how did the outer surface of the brain, the layered cortex of neuronal gray matter, get its folds? Guided by prior experimental observations of the growth of the cortex relative to the underlying white matter, we argue that these folds arise due to a mechanical instability of a soft tissue that grows nonuniformly. Numerical simulations and physical mimics of the constrained growth of the cortex show how compressive mechanical forces sculpt it to form characteristic sulci and gyri, consistent with observations across species in both normal and pathological situations.

APPLIED PHYSICAL SCIENCES

BIOPHYSICS AND COMPUTATIONAL BIOLOGY

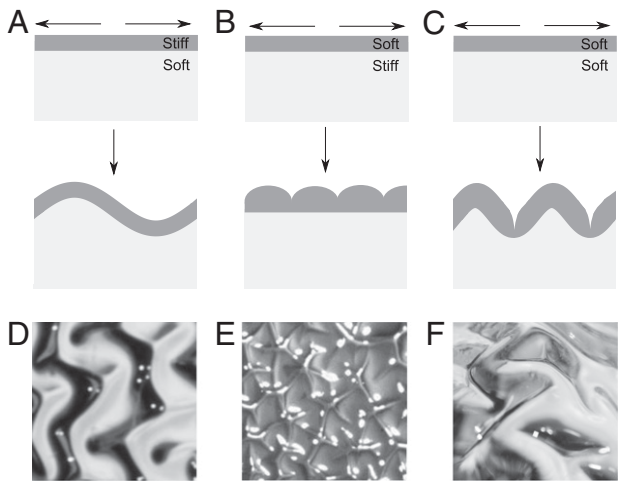

**Fig. 1.** Wrinkling and sulcification in a layered material subject to differential growth. (*A*) If the growing gray matter is much stiffer than the white matter it will wrinkle in a smooth sinusoidal way. (*B*) If the gray matter is much softer than the white matter its surface will invaginate to form cusped folds. (*C*) If the two layers have similar moduli the gray matter will both wrinkle and cusp giving gyri and sulci. Physical realizations of *A*, *B*, and *C*, based on differential swelling of a bilayer gel (*Materials and Methods*), confirm this picture and are shown in *D*, *E*, and *F*, respectively.

instability discussed earlier, and can lead to an emergent pattern very reminiscent of sulci and gyri in the brain.

### Results

A physical experiment to mimic these patterns can be easily created using a hemispherical polydimethylsiloxane (PDMS) gel coated with a layer of PDMS that can swell by absorbing a solvent such as hexanes (*Materials and Methods*). By varying cross-linking densities we can prepare samples with different ratios of the moduli of the two layers and capture both the wrinkled morphology shown in Fig. 1*D* (when the outer layer is stiffer) and the sulcified morphology shown in Fig. 1*E* (when the outer layer is softer). In particular, we see the appearance of brainlike morphologies with deep sulci when the modulus ratio is close to unity (Fig. 1*F*).

To study gyrification quantitatively, we first construct a numerical model in two dimensions. We start with a rectangular domain consisting of a layer of gray matter on top of a deep layer of white matter, both having the same uniform shear modulus $\mu$. The material is assumed to be neo-Hookean with volumetric strain energy density

$$\mathcal{W} = \frac{\mu}{2}\left[\operatorname{Tr}(\mathbf{FF}^T)J^{-2/3} - 3\right] + \frac{K}{2}(J-1)^2, \quad [1]$$

where $\mathbf{F}$ is the deformation gradient, $J = \det(\mathbf{F})$, and the bulk modulus $K = 10^3\mu$ makes the tissues almost incompressible. To model growth of the gray matter relative to the white matter, we apply a tangential growth profile,

$$g(y) = 1 + \frac{\alpha}{1 + e^{10(y/T-1)}}, \quad [2]$$

so that $g = 1$ in the white matter and $g = 1 + \alpha$ in the gray matter, with a smoothed step at the interface (Fig. S1). Here, $y$ is distance from the top surface in material coordinates, $T$ is the undeformed thickness of the gray matter, and $\alpha$ controls the magnitude of expansion. Later on we denote $g \equiv 1 + \alpha \approx g(0)$. We use a custom finite element method to minimize the elastic energy (details Please select the in *Materials and Methods*). Our 2D plane-strain calculations also include constrained expansion in the $z$ direction, although folding can only occur in the $x-y$ plane; we find that when transversely isotropic tangential expansion exceeds $g = g_x = g_z \approx 1.29$ sulcification of the gray matter becomes energetically favorable over a smooth surface, and the gray matter forms cusped folds largely internal to the gray matter and reminiscent of the folds in lightly sulcified brains such as the porcupine (Fig. 2*A*). As $g_x$ is increased further (for simplicity $g_z = 1.29$ was fixed) the gray matter folds down into the white matter forming a big cusped sulcus and smooth gyrus, reminiscent of the sulci and gyri found in more folded brains such as a cat (Fig. 2*B*). Our plots also indicate regions of compressive and tensile stress, which agree with observations in developing ferret brains (11).

We plot the geometric characteristics of the sulcus, such as depth and width, as a function of $g_x$ in Fig. 2*C*, which allow us to establish several nontrivial similarities between our geometry and actual brains (Fig. S2). After the transition from smooth to sulcified, the sulcal depth increases continuously although the width between sulci is always finite, agreeing with observations in weakly convoluted brains. In the high $g$ regime, the optimal spacing is about $4T$ whereas depth continues to increase, in agreement with observations in highly folded brains. Finally, the deformed thickness of the gray matter varies such that, at the gyrus, it is nearly twice that at the base of the sulcus; the same pattern as seen in all real brains. Our 2D model thus captures the essential features of individual sulci and gyri and the intersulcal spacing.

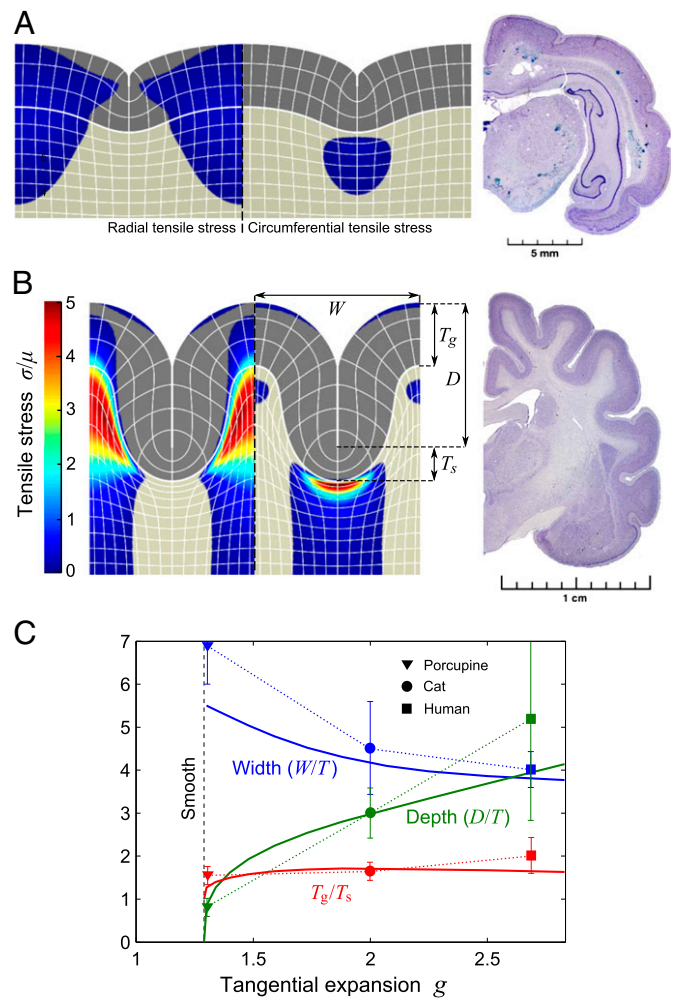

**Fig. 2.** Formation of a minimal sulcus. The 2D sulci with tangential expansion ratio of (*A*) $g = 1.30$ and (*B*) $g = 2.25$ of the gray matter (Eq. 2 and Fig. S1). Coloring shows radial and circumferential tensile stress in the left and right sulci, respectively. The stress is compressive in the noncolored areas. Grid lines correspond to every 20 rows or columns of the numerical discretization with nodes. The width $W$, depth $D$, and thickness of the gray matter in the sulcus ($T_s$) and gyrus ($T_g$) are indicated in *B*. For comparison with observations of brains, we also show sections of porcupine and cat brains, taken from www.brainmuseum.org. (*C*) Scaled dimensions of the simulated sulcus (solid lines) as a function of $g$ compared with those in porcupine (triangles), cat (dots), and human (squares) show that our model can capture the basic observed geometry. Width and depth are given relative to the undeformed thickness $T$ of the gray matter (for details of the measurements and error bars, see Fig. S2).

Although sulci are fundamentally different from wrinkles, a qualitative understanding of our results follows by using the classical formula $\lambda = 2\pi t[\mu/(3\mu_s)]^{1/3}$ for the wrinkling wavelength of a compressed stiff film (modulus $\mu$, thickness $t$) on a soft substrate (modulus $\mu_s$) (27). Extrapolating this to the case here ($\mu_s = \mu$) yields $\lambda \approx 4.36t$ in rough agreement with the simulated sulcal spacing. A rigorous analytical treatment of gyrification is, however, presently out of reach due to the subcritical nature of the instability that is accompanied by finite strains and cusplike features. Although the underlying mechanical principle is that the gray matter folds to relax its compressive stress and that is balanced by deforming the white matter, we emphasize that the details are quite different from wrinkling and buckling, because sulcification is a scale-free nonlinear subcritical instability (24).

We now explore the patterns of sulci and gyri in 3D by modeling the brain as a thick spherical shell, with outer radius $R$ and inner radius $R_i = R/2$, including both the gray and white matter; we note that for such geometries the resulting gyrification patterns are independent of the (presence or absence of a) core. As in the 2D model above, the domain is assumed to be of uniform elastic material described by Eq. **1**, but for numerical convenience we now adopt modest compressibility with $K = 5\mu$, corresponding to Poisson's ratio $\nu \approx 0.4$. Brain tissues actually show time-dependent compressibility owing to poroelasticity (28), but this is irrelevant over the long times associated with morphogenesis, when we may safely limit ourselves to considering just elastic effects. We assume that tangential expansion, given by Eq. **2**, is transversely isotropic so that the area expansion is given by $g^2$. We model small brains as complete thick spherical shells and patches of large brains as patches of thick spherical shells with periodic boundary conditions along the edges, and discretize them using tetrahedral elements (*Materials and Methods*).

Our 3D model of the brain has three geometrical parameters, brain radius (size) $R$, cortical thickness $T$ and the tangential expansion $g^2$ with experimentally observable analogs. Indeed, in mammals the cortical thickness $T$, gray-matter volume $V_G$ and white-matter volume $V_W$ are linked by robust scaling laws that relate brains varying over a millionfold range in weight (29, 30). These laws can be written, in dimensionless units, as $T \sim V_G^{0.1}$ and $V_W \sim V_G^{1.23}$ (29). Using the spherical geometry of our model, we can relate these quantities to our model parameters as $V_G \sim g^2[R^3 - (R - T)^3]$ and $V_W \sim (R - T)^3 - (R/2)^3$. These empirical scaling laws, together with an estimate that $g^2 = 5$ for $R/T = 20$, eliminate two degrees of freedom from the model, leaving us with a single parameter family of models describing brains of different sizes. This is shown in Fig. 3 where we plot $g^2$ against relative brain size $R/T$ and display images of real brains as well as numerically simulated brain shapes for a range of representative relative sizes.

Our model correctly predicts that brains with $R/T \lesssim 5$ (corresponding to physical size of $R \approx 5$ mm) should be smooth as $g$ is insufficient to cause buckling. Intermediate-size brains are correctly predicted to have isolated sulci that are largely localized within the gray matter. Larger brains become increasingly folded, with sulci penetrating the white matter and the brain surface displaying complicated patterns of branched sulci, similar to those in large real brains (Fig. 3). The degree of folding is conventionally quantified by the gyrification index (GI), the ratio of the surface area to the area of the convex hull. For the largest brain that we simulate ($R/T = 20$, $g^2 = 5$, corresponding to physical size $R \approx 36$ mm) we find GI $\approx 2.8$, which can be compared with the modestly larger human brain that has GI $\approx 3$ in regions that exclude the sylvian fissure (31). Most actual brain shapes deviate from spherical so that sulci, especially in small- and medium-sized brains, tend to align with the direction of least curvature. By repeating our calculations on an ellipsoidal geometry (Fig. 3) we capture this qualitative trend. The numerical brain shapes are complemented by experimental realizations in Fig. 3. Our bilayer gel brain models (*Materials and Methods*) capture the realistic sulcal spacing of about $4T$ and the qualitative trends in variation of sulcal patterns with $R/T$ up to modest-size brains.

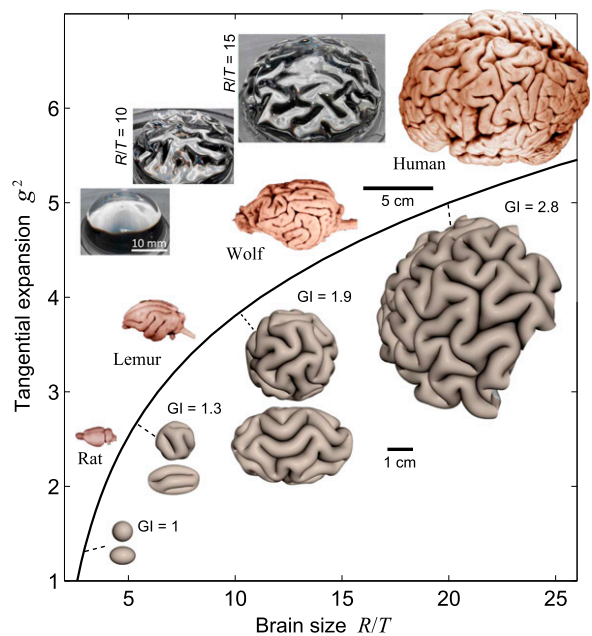

**Fig. 3.** Known empirical scaling laws for gray-matter volume and thickness are mapped on a $g^2$ vs. $R/T$ diagram. Corresponding simulations for spherical brain configurations, with images shown at a few points, show that the surface remains smooth for the smallest brains, but becomes increasingly folded as the brain size increases. We also show patterns for ellipsoidal configurations (major axis = 1.5 × minor axes) that lead to anisotropic gyrification. Images of rat, lemur, wolf, and human brains illustrate the increasingly prominent folding with increasing size in real brains. Also shown are images of our physical mimic of the brain using a swelling bilayer gel of PDMS immersed in hexanes. The smooth initial state gives rise to gyrified states for different relative sizes of the brain $R/T = 10, 15$ (see also Fig. 5). All of the brain images are from www.brainmuseum.org.

Because many brain atlases show different sections of the brain to highlight the anatomical complexity of the folds, we show sections of our simulated patterns in Fig. 4A. For comparison, we show sections from a raccoon brain, which has a similar size to the simulated brain, and see that the two appear very similar. An important observation that becomes apparent is that cutting through gyri and sulci with various alignments with respect to the section plane gives the impression of rather complex gyrification and exaggerates depths of sulci, especially when the section plane is off the center of curvature (plane 2; Fig. 4A). Simulated cross sections display features such as buried gyri and regions with disproportionally thick cortices seen in sections of real brains, but they are really just geometric artifacts of sectioning.

As can be observed, the calculated gyri are rounded rather than flattened as in, e.g., brain samples that have been fixed before removal. Experimental evidence suggests, in agreement with our model, that the compressive constraint of the skull or meninges is not required for gyrification (30), but it could affect the appearance by flattening the gyral crowns. Our simulation of a brain confined by a rigid shell that mimics the skull confirms this (Fig. 4B).

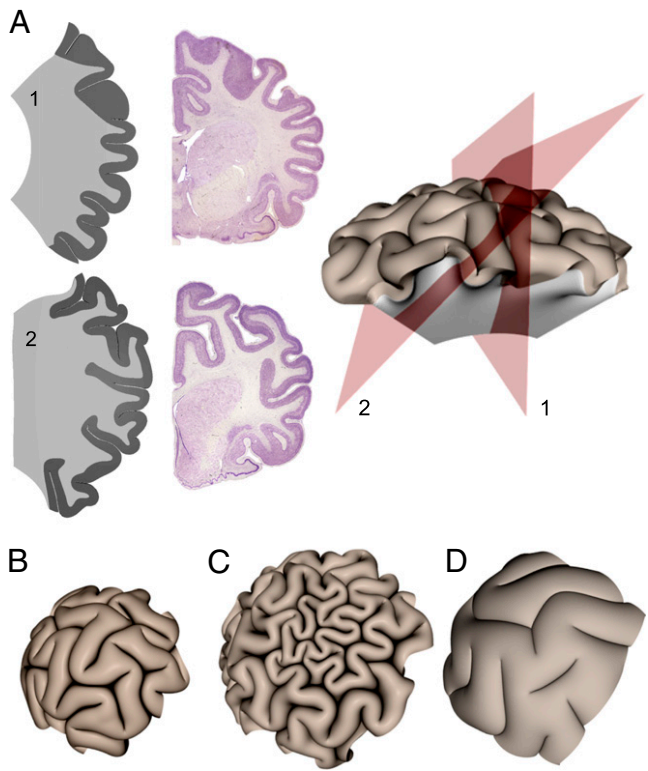

**Fig. 4.** (*A*) Sections of a simulated brain (section planes indicated at right) are compared with coronal sections of a raccoon brain (from www.brain-museum.org). Cuts through the center of the brain (*Upper*) and the off center (*Lower*) show that we can capture the hierarchical folds but emphasize how misleading sections can be in characterizing the sulcal architecture. (*B*) Confining our simulations with a uniform pressure of $0.7\mu$ to mimic the meninges and skull leads to a familiar flattened sulcal morphology. (*C*) Changing the gray matter thickness in a small patch of the growing cortex leads to morphologies similar to polymicrogyria in our simulations. Here $g^2 = 5$ and $R/T = 20$ except in the densely folded region where $R/T = 40$. (*D*) A simulated brain of same physical size as that in C but with a thickened cortex ($R/T = 12$) and reduced tangential expansion ($g^2 = 2$) displays wide gyri and shallow sulci reminiscent of pachygyric brains.

The empirical scaling laws we used to restrict the parameter space of our model describe normal brains. Deviations from the scaling laws relating gyrification index to the relative cortical thickness are known to lead to some pathological morphologies in brains. Our model can reproduce these variations by considering deviations in the parameters $R/T$ and $g$. For example, polymicrogyria is typically associated with the gray matter having only four layers rather than six (32), and hence being thinner. In the context of our model, reducing $T$ while maintaining $g$ for a fixed brain size results in more and smaller sulci and gyri, as shown in Fig. 4*C*, consistent with observations. Pachygyria and lissencephalia are associated with reduced neuronal migration to the cortex and thus less surface expansion leading to smaller values of $g$ (33). In the context of our model, smaller surface expansion $g$ or a thicker cortex $T$ leads to fewer sulci and wider gyri, as shown in the simulations in Fig. 4*D*, consistent with observations.

## Discussion

Our mechanical model for gyrification based on biological observations and physical constraints uses realistic but simplified geometries, growth profiles, and mechanical properties. Real brain tissues are known to change their growth rates in response to stress (20), and other mechanisms lead to relaxation of elastic stresses over time. Nevertheless, our simple elastic description suffices to explain the essentials of gyrification while more complicated inelastic material behavior has a minor role. Our theory naturally leads to cusped sulci and smooth gyri, and shows that gray-matter thickness should vary by about a factor of two between gyral crowns and sulcal fundi, and that sulcal depths should vary substantially relative to gray-matter thickness whereas gyral spacings should be always about 4 times the gray-matter thickness; these are consistent with numerous experimental observations (12, 30). Furthermore, the model captures the variations in the patterns of sulci as a function of brain size showing that small brains should be smooth and large brains should have complex branched sulcal networks, in line with observations. Finally, our model predicts that gray matter is under compression and white matter is under radial tension beneath gyri and circumferential tension beneath sulci, also consistent with recent observations (11). From a broader perspective, our study complements other recent studies that highlight the role of mechanical forces on morphogenesis in such instances as gut patterning (34) where patterns similar to gyri and sulci yield villi in the gut lumen.

The shape of the mammalian brain has a number of prominent large-scale features: the longitudinal fissure, the sylvian fissure, and many specific gyri and sulci with important functional correlates that we cannot capture with our minimal model. Thus, our approach should be seen as complementing the detailed variations in proliferation patterns, cell migration, shape change, etc., that give rise to cortical expansion and in turn are regulated by it. Indeed, the growing insight into regulatory mechanisms of neuronal proliferation and migration (3, 4, 5, 6) implicates genetic mutations responsible for brain size (e.g., ASPM, CENPJ for microcephaly), cortex thickness (e.g., GPR56 for polymicrogyria), and the gyrification index (e.g., RELN for lissencephaly), and thus links naturally to the parameters in our model for brain morphology. Simultaneously, comparative phylogenies that look at expression profiles for these genes (6) may also provide an evolutionary window into brain gyrification, wherein proximal genetic causes lead to pathways that allowed for the growth of the brain and the expansion of the cortex, turning primitive smooth brains to current gyrified brains, leaving open the possibility that the process is now significantly regulated. Finally, our geometric mechanics approach to gyrification might also serve as a template for brain morphometrics in terms of experimentally measurable geometrical parameters that vary across individuals, complementing current geometric statistics approaches (35) by accounting for the underlying physical processes.

## Materials and Methods

**Physical Models.** To create a simulacrum of the brain, we used a poly(dimethylsiloxane) (PDMS) elastomer bilayer consisting of an outer layer coated on the surface of a hemispherical core. Differential swelling of the top layer when exposed to a solvent (hexanes) mimics constrained cortical expansion and leads to a variety of surface morphologies. The hemispherical core is fabricated with PDMS elastomer (Sylgard 184; Dow Chemical Co., Midland, MI) using replica molding, using a 30:1 mass ratio of base monomer to cross linker that is mixed and poured over a concave lens-shaped silicone mold with $r \approx 11$ mm. The mixture is left at room temperature to allow trapped air bubbles to escape and then cured at 75 °C for about 30 min to produce a partially cured contact surface. After cooling, the PDMS core is carefully peeled off from the mold. The partially cured PDMS surface is sticky and thus is used to enhance the interfacial bonding between the top layer and the core for the preparation of a bilayer specimen; $\mu_{0c} \approx 100$ kPa in the fully cured state (36). The specimen geometry and the experimental setup are illustrated in Fig. 5.

The top layers are prepared with different compositions of the PDMS mixture by a drop coating technique, in which layer thickness is controlled by adjusting the amount and viscosity of uncured PDMS mixture deposited on the hemispherical core in a preheated oven. The selected elastomer composition is first thoroughly mixed and degassed to eliminate air bubbles. Then, drops of a mixture are deposited and uniformly spread onto the surface of a preprepared PDMS core using a pipette. Finally, the specimen is completely cured at 75 °C overnight. The thickness of the prepared top layers $T_0$ is determined by comparing the optical images before and after layer coating using an image analysis program.

The prepared bilayer specimen is fully immersed in a Petri dish filled with hexanes (Sigma-Aldrich) at room temperature for a period $t$ typically less than

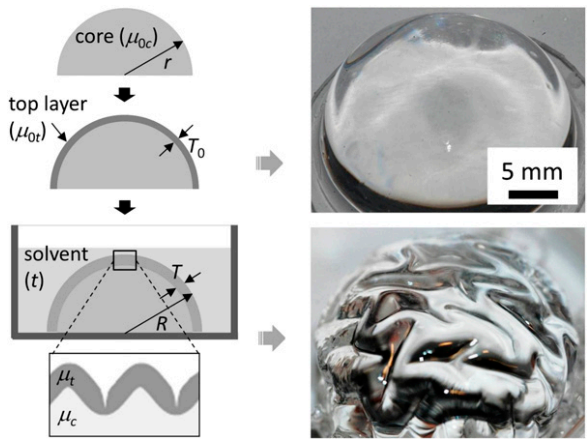

**Fig. 5.** A physical model of brainlike instability. To mimic the growth of the gray matter in the brain, a hemispherical elastomer (radius $r$, shear modulus $\mu_{0c}$) is coated with a top elastomer layer (thickness $T_0$, shear modulus $\mu_{0t}$) that swells by absorbing solvent over time $t$. Representative images of a bilayer specimen in the initial (dried) state and swollen state (modulus ratio $\mu_t/\mu_c \approx 1$) are shown at right.

10 min. This exposure time is short enough to prevent the solvent from penetrating into the core, but is sufficient to swell the top layer. Immediately after solvent exposure, the bilayer specimen is withdrawn from the bath and subsequently imaged with a digital camera (Nikon D80) equipped with a zoom lens (Sigma 105 mm f/2.8 EX DG Macro). Less polar solvents such as chloroform and toluene (both having a similar high PDMS swelling coefficient; ref. 37) yield similar results, but hexanes has a relatively low evaporation rate, which makes the deswelling process slow enough to capture the surface patterns.

The shear modulus of the outer layer in the swollen state $\mu_t$ is lower than that in the initial dry state ($\mu_{0t}$), but the modulus of the core remains unchanged over the short period of solvent exposure time ($\mu_c \approx \mu_{0c}$). We consider three classes of models, sketched in Fig. 1 A–C, where $\mu_t/\mu_c > 1$, $\mu_t/\mu_c < 1$, and $\mu_t/\mu_c \approx 1$. If $\mu_{0t} \gg \mu_{0c}$, even the swollen modulus $\mu_t$ of the top layer remains considerably greater than $\mu_c$, so that $\mu_t/\mu_c > 1$. If however $\mu_{0t} \approx \mu_{0c}$, after swelling of the top layer $\mu_t/\mu_c < 1$. An intermediate situation is when the initial $\mu_{0t}$ is slightly greater than $\mu_{0c}$. If this bilayer is immersed in a solvent, then the modulus of the top layer in the swollen state is comparable to that of the core, i.e., $\mu_t/\mu_c \approx 1$ as in the brain. The above three situations are experimentally tested by varying the modulus of the outer layer ($\mu_{0t} \approx 1$ MPa, $\approx 500$ kPa, and $\approx 100$ kPa, monomer to cross-linker ratios 5:1, 10:1, and 30:1, respectively; ref. 36) while keeping the core modulus constant ($\mu_{0c} \approx 100$ kPa). The resulting instability patterns are shown in Fig. 1 D–F.

We also experimentally examine the effects of varying the thickness of the top layer $T_0$ while keeping all other properties fixed ($\mu_{0t} \approx 500$ kPa, $\mu_{0c} \approx 100$ kPa, and $r \approx 11$ mm) by using a set of nine bilayer specimens with top layers of three different thicknesses, namely $\approx 300$ $\mu$m, $\approx 800$ $\mu$m, and $\approx 1.2$ mm (hence, the relative initial radius $R/T_0 \approx 38$, 15, and 10, respectively, where $R = r + T_0$). Each specimen is immersed in hexanes for a specific period $t$, and the resulting surface morphologies induced by differential growth and mechanical buckling are imaged (Fig. S3). The absorption of solvent is a diffusive process, with $T = \sqrt{4Dt}$, where $T$ is the penetration depth (or thickness) of solvent into a top layer and $D$ is the diffusion coefficient. Assuming $D \approx 6 \times 10^{-10}$ m$^2$/s for hexanes in PDMS (38, 39), the estimated $T$ (and $R/T$) for $t = 1$ min, $t = 4$ min, and $t = 9$ min is found to be $\approx 380$ $\mu$m (30), 760 $\mu$m (15), and 1.1 mm (10), respectively. Highlighted images along a diagonal line in Fig. S3 show the case where the initial radius $R/T_0$ closely matches the estimated relative radius $R/T$. Two of these images are compared with the corresponding images of real and simulated brain shapes in Fig. 3.

**Numerical Simulations.** Our finite element model for brain folding is based on constant strain triangle (2D) or tetrahedron (3D) elements and an explicit solver for quasistatic equilibration of the system and allows simulation of the large strains and highly nonlinear mechanics involved in gyrification, but necessitates the use of high density meshes.

In 2D simulations the domain is discretized into a rectangular lattice of 150 × 600 nodes (width × depth) and filled with plane-strain triangle elements forming a mesh of crossed triangles. The top surface of the gray matter is free and we apply symmetric boundary conditions along the lateral sides of the domain. Owing to symmetry, the simulation domain contains only one half of the sulcus/gyrus. The domain is 10 times the gray-matter thickness (i.e., 60 topmost rows of elements are included in the gray matter), to minimize the effect of substrate thickness on the sulcus and gyrus. At each simulated value of tangential expansion the sulcus is initiated by applying a downward force to the surface node of a lateral boundary, and self-contact of the sulcus is accounted for by preventing the surface crossing the vertical line that defines the boundary. The initiating force is then removed, the system allowed to relax, and the aspect ratio of the domain is adjusted quasistatically (by sweeping the height of the domain over a finite interval enclosing the energy minimum) to find the energetically optimal relative width of the gyrus.

The 3D simulations are based either on irregular tetrahedral meshes (small brain simulations that implement a full spherical or ellipsoidal thick shell) or a curved cubical mesh where each cube is divided to five tetrahedrons (large brain simulations that implement a patch of a spherical thick shell), see Fig. 6. Inner surfaces of the thick shells are clamped, but the use of free boundary conditions or simulations with full solid spheres would yield similar results. The irregular 3D meshes consist of $\approx 2 \times 10^6$ nodes (about $10^7$ elements) and the regular meshes consist of $320 \times 320 \times 80$ nodes (about $4 \times 10^7$ elements). In each case the mesh density is such that the gray-matter layer contains at least eight layers of elements through its thickness. The spherical patch spans an angle of $\pi/2$ about the $x$ and $z$ axes, with periodic boundary conditions so that if a copy of the domain is rotated by an angle $\pi/2$ about the $x$ or $z$ axis, it would connect seamlessly to the original domain. The simulations based on regular meshes have small random spatial variations in growth to break the otherwise perfect rotational symmetry. These random perturbations do not affect the shape or size of sulci and gyri or qualitative features of the sulcal pattern, but different perturbation fields produce different patterns because they are the only mechanism breaking the symmetry in the system. In the simulations based on irregular meshes, the mesh provides sufficient randomness.

A stress-free initial configuration of a tetrahedron is defined by the matrix $\hat{\mathbf{A}} = \begin{bmatrix} \hat{x}_1 & \hat{x}_2 & \hat{x}_3 \end{bmatrix}$, where $\hat{x}_1$, $\hat{x}_2$ and $\hat{x}_3$ are vectors describing the tetrahedron, assuming a Cartesian coordinate system. The deformed configuration of the tetrahedron, including growth and elastic deformation, is defined by

$$\mathbf{A} = \begin{bmatrix} x_1 & x_2 & x_3 \end{bmatrix} = \mathbf{FG}\hat{\mathbf{A}}, \quad [3]$$

where $x_1$, $x_2$, and $x_3$ are the deformed basis vectors and $\mathbf{F}$ is the elastic deformation gradient. The growth tensor

$$\mathbf{G} = g\mathbf{I} + (1-g)\hat{n}_s \otimes \hat{n}_s \quad [4]$$

describes tangential expansion perpendicular to the surface normal $\hat{n}_s$, with $g$ given by Eq. 2. At each time step we obtain $\mathbf{F}$ from Eq. 3 by using $\mathbf{F} = \mathbf{A}(\mathbf{G}\hat{\mathbf{A}})^{-1}$. The Cauchy stress, i.e., the force per unit area in the deformed configuration, is derived from the strain energy density $\mathcal{W}(\mathbf{F})$ (Eq. 1) by

$$\sigma = \frac{1}{J}\frac{\partial \mathcal{W}}{\partial \mathbf{F}}\mathbf{F}^T, \quad [5]$$

where $J = \det(\mathbf{F})$. Surface traction of each deformed face ($i = 1, 2, 3, 4$) of the tetrahedron is given by $s_i = -\sigma n_i$, where $n_i$ are normals with lengths proportional to the deformed areas of the faces. Nodal forces are obtained by distributing the traction of each face equally for its three vertices.

Self-avoidance of the surface is implemented by preventing nodes penetrating element faces at the surface. If a separation $d$ between a node and face is less than the contact offset $h$ (we use $h = a/3$, where $a$ is the mesh spacing in the initial configuration) it is considered a contact and penalized by an energy $4Ka^2\left(\frac{h-d}{h}\right)^2$. The contact force from this potential is interpolated to the vertex nodes of the face and an opposite force is given to the node in contact with the face.

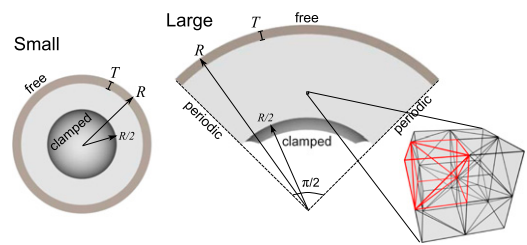

**Fig. 6.** Cross-section views of 3D simulation geometries for small and large brains in their initial undeformed states. The gray-matter thickness $T$, brain radius $R$, and boundary conditions are indicated. A detailed image of the regular mesh structure of the large brain domain shows the reflection symmetry between every pair of elementary cubes that share a face.

Tallinen et al.  PNAS | September 2, 2014 | vol. 111 | no. 35 | 12671

The energy of the system is minimized by damped second-order dynamics, using an explicit scheme,

$$\mathbf{v}(t+\Delta t) = \mathbf{v}(t) + \frac{\mathbf{f}(t) - \gamma \mathbf{v}(t)}{m} \Delta t, \quad [6]$$

$$\mathbf{x}(t+\Delta t) = \mathbf{x}(t) + \mathbf{v}(t+\Delta t) \Delta t. \quad [7]$$

Here $\Delta t = 0.05 a/\sqrt{K}$ is the time step, $m = a^3$ mass of a node, and $\gamma = 10\,m$ viscous damping. Vectors $\mathbf{f}$, $\mathbf{v}$, and $\mathbf{x}$ are force, velocity, and position of a node, respectively. The 2D simulations are implemented similarly but with triangular elements instead of tetrahedra.

**ACKNOWLEDGMENTS.** We thank Teijo Kuopio for discussions. Computations were run at CSC–IT Center for Science, Finland. The brain images were obtained from the collections of the University of Wisconsin, Michigan State, and the National Museum of Health and Medicine, funded by the National Science Foundation and the National Institutes of Health. We thank the Academy of Finland (T.T.), the Wyss Institute for Biologically Inspired Engineering (J.Y.C. and L.M.), the Royal Society (J.S.B.), the Singleton Foundation (L.M.), and the MacArthur Foundation (L.M.) for partial financial support.

# Supporting Information

## Tallinen et al. 10.1073/pnas.1406015111

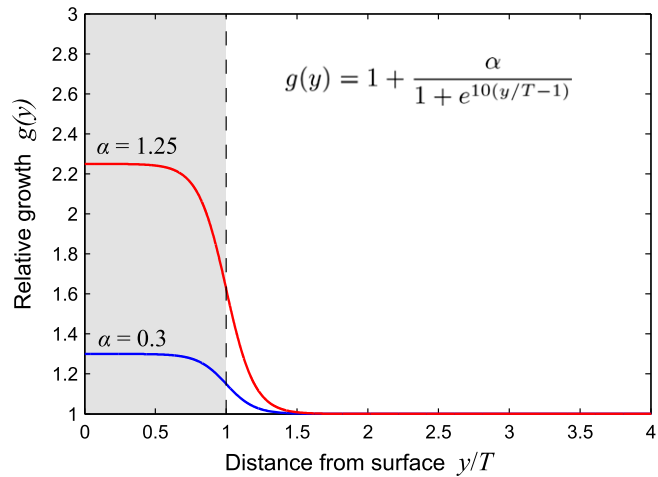

Fig. S1. Tangential growth profiles for $\alpha = 0.3$ and $\alpha = 1.25$ applied in simulations of Fig. 2 *A* and *B*, respectively.

$$g(y) = 1 + \frac{\alpha}{1 + e^{10(y/T - 1)}}$$

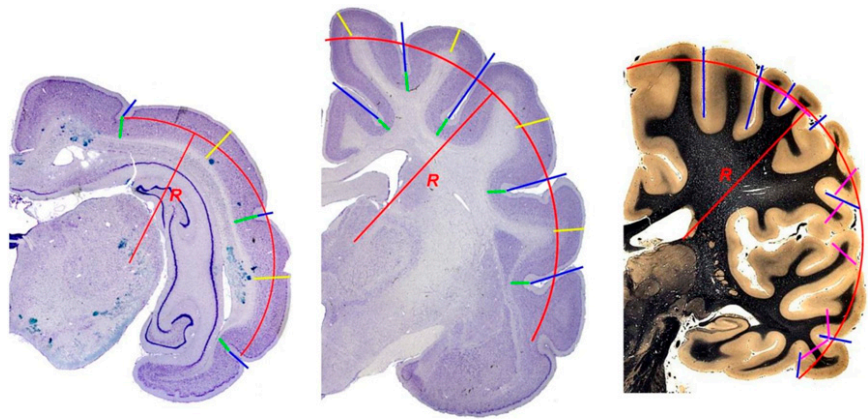

|  |  | Porcupine | Cat | Human |
|---|---|---|---|---|
| Radius | $R$ [mm] | 6.5 | 9.2 | 52 |
| Gray matter thickness | $T$ [mm] | 1.1 | 0.9 | 2.1 |
| Tangential expansion | $g$ | 1.3 | 2.0 | 2.7 |
| Radius (normalized) | $R/T$ | 5.9 | 10.2 | 25 |
| Gyral width | $W/T$ | 6.9 ± 0.9 | 4.5 ± 1.1 | 4.0 ± 0.4 |
| Sulcal depth | $D/T$ | 0.8 ± 0.2 | 3.0 ± 0.6 | 5.2 ± 2.5 |
| GM thickness ratio | $T_g/T_s$ | 1.6 ± 0.2 | 1.6 ± 0.2 | 2.0 ± 0.4 |

Fig. S2. Geometric parameters from brain sections of a porcupine, cat, and human. Brain radius $R$ is indicated by the red arcs. Gyral widths in the porcupine and cat are determined as the length of the red arc over each gyrus. In the human the sulcal geometry is more complicated and some gyri are inclined with respect to the sectioning plane. Therefore, in the human gyral widths are determined more selectively as indicated by magenta line segments. Sulcal depths are indicated by blue line segments (the sylvian fissure and sulci that are clearly inclined with respect to the sectioning plane are excluded in the human). The thickness of the gray matter at the gyri is indicated by the yellow line segments, and thickness of the gray matter at the sulci by the green line segments (not shown for the human). The undeformed thickness of the gray matter is approximated by $T = T_g/1.5$ using the mean thickness $T_g$ of the gray matter at the gyri. Tangential expansion $g$ is estimated by dividing the length of the surface contour by the length of the red arc (excluding the sylvian fissure in the human). The data shown for $W/T$, $D/T$, and $T_g/T_s$ are given as the mean ± SD. All images are cell-stained (porcupine and cat) or fiber-stained (human) coronal sections from www.brainmuseum.org.

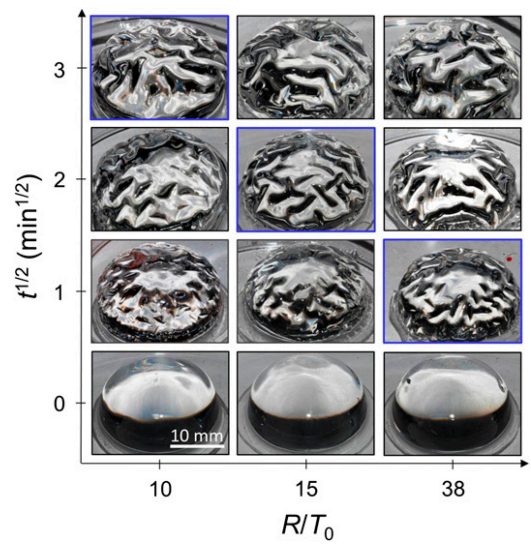

**Fig. S3.** For the swelling gel model that mimics the growing brain, we show the sulcal patterns as a function of the scaled initial radius $R/T_0$ and solvent exposure time $t$. The sulcal spacing of the patterns increases with decreasing $R/T_0$ and increasing $t$ in qualitative agreement with our theoretical predictions. The images highlighted with a blue border represent states where the solvent penetration depth is comparable with the upper layer thickness.